\documentclass[english,aip,apl,superscriptaddress,reprint]{revtex4-1}
\usepackage{mathptmx}
\usepackage[T1]{fontenc}
\usepackage[latin9]{inputenc}
\usepackage{verbatim}
\usepackage{amsmath}
\usepackage{graphicx}

\makeatletter
\@ifundefined{textcolor}{}
{%
 \definecolor{BLACK}{gray}{0}
 \definecolor{WHITE}{gray}{1}
 \definecolor{RED}{rgb}{1,0,0}
 \definecolor{GREEN}{rgb}{0,1,0}
 \definecolor{BLUE}{rgb}{0,0,1}
 \definecolor{CYAN}{cmyk}{1,0,0,0}
 \definecolor{MAGENTA}{cmyk}{0,1,0,0}
 \definecolor{YELLOW}{cmyk}{0,0,1,0}
 }

\usepackage[titles]{tocloft}
\renewcommand\cftdotsep\cftnodots
\cftsetindents{figure}{0em}{1.5em}
\cftpagenumbersoff{figure}

\makeatother

\usepackage{babel}

\begin{document}

\title{Plasmonic rod dimers as elementary planar chiral meta-atoms}

\author{Sergei V. Zhukovsky}

\affiliation{Institute of High-Frequency and Communication Technology, Faculty
of Electrical, Information and Media Engineering, University of Wuppertal,
Rainer-Gruenter-Str. 21, D-42119 Wuppertal, Germany}

\affiliation{Department of Physics and Institute for Optical Sciences, University
of Toronto, 60 St. George Street, Toronto, Ontario M5S 1A7, Canada}

\author{Christian Kremers}

\affiliation{Institute of High-Frequency and Communication Technology, Faculty
of Electrical, Information and Media Engineering, University of Wuppertal,
Rainer-Gruenter-Str. 21, D-42119 Wuppertal, Germany}

\author{Dmitry N. Chigrin}

\affiliation{Institute of High-Frequency and Communication Technology, Faculty
of Electrical, Information and Media Engineering, University of Wuppertal,
Rainer-Gruenter-Str. 21, D-42119 Wuppertal, Germany}
\begin{abstract}
Electromagnetic response of metallic rod dimers is theoretically calculated
for arbitrary planar arrangement of rods in the dimer. It is shown
that dimers without an in-plane symmetry axis exhibit elliptical dichroism
and act as \textquotedblleft{}atoms\textquotedblright{} in planar
chiral metamaterials. Due to a very simple geometry of the rod dimer,
such planar metamaterials are much easier in fabrication than conventional
split-ring or gammadion-type structures, and lend themselves to a
simple analytical treatment based on coupled dipole model. Dependencies
of metamaterial's directional asymmetry on the dimer's geometry are
established analytically and confirmed in numerical simulations.
\end{abstract}


\maketitle
\begin{figure}
\centering{}\includegraphics[width=0.92\columnwidth]{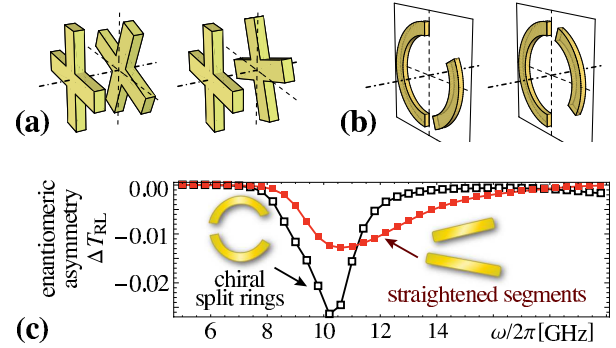}\caption{(Color online) Example of (a) 3D and (b) 2D enantiomeric meta-atoms;
(c) difference in transmission for LH/RH-polarized wave for split-ring
vs. {}``straighened split-ring'' (rod dimer) structure. \label{fig:intro}}
\end{figure}

Metamaterials have attracted intense scientific interest in recent
years for their unusual physical properties rare or absent in nature.
One example of such properties is giant optical activity \cite{1-Giant05}
in composite materials containing spiral-like or otherwise twisted
elements ({}``meta-atoms''). Recently, planar chiral metamaterials
(PCMs) were also introduced \cite{2-Fedotov06}. In PCMs, the meta-atoms
possess two-dimensional (2D) rather than three-dimensional (3D) enantiomeric
asymmetry (Fig.~\ref{fig:intro}a--b). PCMs are distinct from both
3D chiral and Faraday media in that their polarization eigenstates
are co-rotating elliptical rather than counter-rotating elliptical
or circular \cite{2-Fedotov06}. This leads to exotic polarization
properties, e.g., asymmetry in transmission for left-handed (LH) vs.
right-handed (RH) circularly polarized incident wave without nonreciprocity
present in Faraday media. Such exotic properties, combined with the
small dimensions of planar structures, make PCMs promising for polarization
sensitive integrated optics applications.

Several different PCM designs have been proposed so far, the most
notable example being the asymmetric chiral split rings \cite{3-Plum09}.
Still, this geometry is complicated enough to make microscopic theoretical
analysis difficult, and it can be shown that simpler geometries may
be sufficient to achieve the desired chiral properties. Indeed, simulations
reveal (Fig.~\ref{fig:intro}c) that straightening the split-ring
segments into two rods (preserving the segment's length and mutual
orientation) results in similar transmission asymmetry for LH vs.
RH circularly polarized light. Rods, unlike more complicated particles,
lend themselves to a quite straightforward theoretical analysis (see,
e.g., \cite{LedererPRB08}). Hence it is of interest to explore the
potential of two-rod dimers as meta-atoms for PCMs. In addition, rod-like
meta-atoms are much easier for fabrication than split-ring or gammadion
structures, especially in the optical domain.

In this Letter, we show that double-rod plasmonic dimers can function
as planar chiral meta-atoms. We find that the effective permittivity
of a planar rod dimer supports elliptical dicroism necessary for PCM
effects \cite{4-Zhukovsky09}. We confirm both analytically and numerically
that any dimer with distinct 2D enantiomers intrinsically exhibits
planar chiral behavior, and systematically investigate the relations
between the strength of chiral properties and the dimer\textquoteright{}s
geometrical parameters.\textbf{ }By doing so, we propose a simple
PCM design that lends itself to easy fabrication and simple analytical
treatment.

\begin{figure}
\centering{}\includegraphics[width=1\columnwidth]{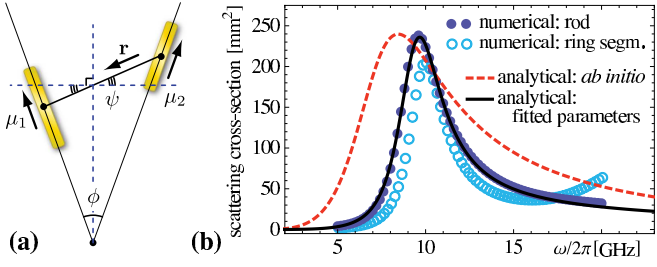}\caption{(Color online) (a) Geometry of an arbitrary planar rod dimer considered
in the present study. (b) Scattering cross-section of a parallelepiped-shaped
rod with dimensions $13\times0.8\times0.8$ mm determined by numerical
simulations and using Eq.~\eqref{eq:eq1},\emph{ ab initio} and using
best-fit parameters. Numerical results for a 13 mm long ring segment
are included for comparison.\label{fig:scheme}}
\end{figure}

We consider the structure shown in Fig.~\ref{fig:scheme}a. Since
both rods in the dimer are elongated, it is reasonable to start with
the assumption that they respond to the external electric field $\mathbf{E}_{0}$
with an induced dipole moment $\mathbf{d}_{1,2}=\widehat{\alpha}_{1,2}\mathbf{E}_{0}$
where $\widehat{\alpha}_{j}$ is the polarizability tensor of a single
rod. For a parallelepiped-shaped rod one can follow the depolarization
field approach \cite{5-Moroz09} to obtain polarizability in the form\begin{equation}
\widehat{\alpha}_{j}=\frac{f_{j}\vec{\mu}_{j}\otimes\vec{\mu}_{j}}{\omega_{j}^{2}-\omega^{2}-i\omega\left(\gamma_{j}+\delta_{j}\omega^{2}\right)}\equiv\alpha_{j}\vec{\mu}_{j}\otimes\vec{\mu}_{j}\label{eq:eq1}\end{equation}
where the unit vector $\vec{\mu}_{j}$ denotes the rod orientation;
$f_{j}$, $\omega_{j}$, $\gamma_{j}$, and $\delta_{j}$ are geometry-dependent
parameters, and $\otimes$ stands for tensor (dyadic) product. Fig.~\ref{fig:scheme}b
shows that Eq.~\eqref{eq:eq1} with parameters directly obtained
from the rod dimensions and material properties provides a decent
reproduction for the scattering cross-section spectrum of the rod.
By letting the parameters $f_{j}$, $\omega_{j}$, $\gamma_{j}$,
and $\delta_{j}$ be adjusted, a nearly perfect best-fit between the
expression \eqref{eq:eq1} and numerical simulations can be achieved.
Also note that when a rod is bent to represent a ring segment, its
scattering cross-section near its fundamental resonance does not undergo
significant changes. 

When there are two rods in a dimer, the external field $\mathbf{E}_{0}$
should be modified due to the presence of the neighboring rod. This
results in a system of coupled equations for the dipole moments of
rods \begin{equation}
\mathbf{d}_{1,2}=\widehat{\alpha}_{1,2}\left[\mathbf{E}_{0}+\frac{k^{2}}{\epsilon_{0}}\widehat{\mathbf{G}}(\mathbf{R})\mathbf{d}_{2,1}\right]\label{eq:eq2}\end{equation}
where $\mathbf{R}=R\mathbf{r}$ is the vector connecting the rods
(Fig.~\ref{fig:scheme}a) and $\widehat{\mathbf{G}}$ is the dyadic
Green\textquoteright{}s function in free space:\begin{equation}
\begin{gathered}\widehat{\mathbf{G}}(\mathbf{R})=\frac{e^{ikR}}{4\pi R}\left[\left(1+\frac{ikR-1}{k^{2}R^{2}}\right)\widehat{\mathrm{I}}+\frac{3-3ikR-k^{2}R^{2}}{k^{2}R^{2}}\frac{\mathbf{R}\otimes\mathbf{R}}{R^{2}}\right]\\
\equiv G_{I}\widehat{\mathrm{I}}+G_{R}\mathbf{r}\otimes\mathbf{r}.\end{gathered}
\label{eq:eq3}\end{equation}
with $\widehat{\mathrm{I}}$ being the identity tensor. Solving Eqs.~\eqref{eq:eq2}
with \eqref{eq:eq3} for $\mathbf{d}_{1,2}$, one arrives at the total
dipole moment $\mathbf{d}=\mathbf{d}_{1}+\mathbf{d}_{2}=\widehat{\alpha}_{\text{eff}}\mathbf{E}_{0}$
of the dimer with effective polarizability tensor\begin{equation}
\begin{gathered}\widehat{\alpha}_{\text{eff}}=\frac{\alpha_{1}\vec{\mu}_{1}\otimes\vec{\mu}_{1}+\alpha_{2}\vec{\mu}_{2}\otimes\vec{\mu}_{2}+\alpha_{1}\alpha_{2}\kappa\left(\vec{\mu}_{1}\otimes\vec{\mu}_{2}+\vec{\mu}_{2}\otimes\vec{\mu}_{1}\right)}{1-\alpha_{1}\alpha_{2}\kappa^{2}}\\
\equiv\alpha_{1}^{\text{eff}}\vec{\mu}_{1}\otimes\vec{\mu}_{1}+\alpha_{2}^{\text{eff}}\vec{\mu}_{2}\otimes\vec{\mu}_{2}+\alpha_{3}^{\text{eff}}\left(\vec{\mu}_{1}\otimes\vec{\mu}_{2}+\vec{\mu}_{2}\otimes\vec{\mu}_{1}\right).\end{gathered}
\label{eq:eq4}\end{equation}
where $\kappa$ defines a coupling coefficient between the rods: \begin{equation}
\kappa=\frac{k^{2}}{\epsilon_{0}}\left[G_{I}\left(\vec{\mu}_{1}\cdot\vec{\mu}_{2}\right)+G_{R}\left(\vec{\mu}_{1}\cdot\mathbf{r}\right)\left(\vec{\mu}_{2}\cdot\mathbf{r}\right)\right].\label{eq:kappa}\end{equation}

The effective permittivity tensor $\widehat{\varepsilon}_{\text{eff}}$
is derived from $\widehat{\alpha}_{\text{eff}}$ in Eq.~\eqref{eq:eq4}\textbf{
}as $\widehat{\varepsilon}_{\text{eff}}=\widehat{\mathrm{I}}-(\epsilon_{0}V_{\text{cell}})^{-1}\widehat{\alpha}_{\text{eff}}$.
In axial representation, $\widehat{\varepsilon}_{\text{eff}}$ can
be expressed as \begin{equation}
\widehat{\varepsilon}_{\text{eff}}=\widehat{\mathrm{I}}+\frac{1}{\varepsilon_{0}V_{\mathrm{cell}}}\frac{\alpha_{1}^{\mathrm{eff}}}{2}\left(\mathbf{c}_{+}\otimes\mathbf{c}_{-}+\mathbf{c}_{-}\otimes\mathbf{c}_{+}\right)\label{eq:epsilon_eff}\end{equation}
with the complex vectors $\mathbf{c}_{\pm}=\left(\vec{\mu}_{1}+\eta_{\pm}\vec{\mu}_{2}\right)$
determining optical axes and $\eta_{\pm}$ given by \begin{equation}
\eta_{\pm}=\frac{\alpha_{3}^{\mathrm{eff}}\pm\sqrt{\left(\alpha_{3}^{\mathrm{eff}}\right)^{2}-\alpha_{1}^{\mathrm{eff}}\alpha_{2}^{\mathrm{eff}}}}{\alpha_{1}^{\mathrm{eff}}}.\label{eq:eta12}\end{equation}

\begin{figure}
\centering{}\includegraphics[width=1\columnwidth]{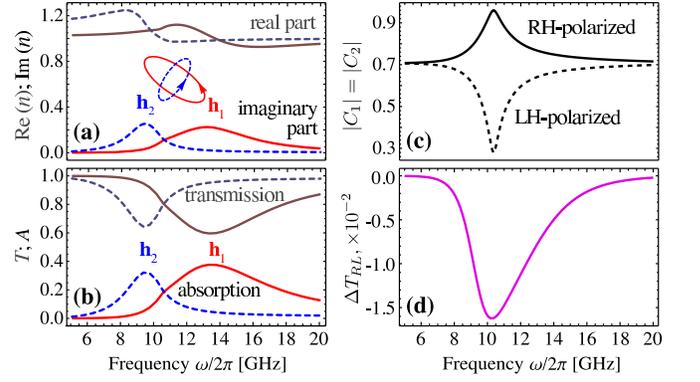}\caption{(Color online) (a) Refractive indices for two co-rotating elliptically
polarized eigenwaves {[}eigenvectors $\mathbf{h}_{1,2}$ of the tensor
$\widehat{N}_{H}$ in Eq.~\eqref{eq:refr_tensor}{]} in the effective
medium given by Eq.~\eqref{eq:epsilon_eff}. (b) Transmission and
absorption for incident wave with polarization coincident with $\mathbf{h}_{1,2}$.
(c) Absolute value of the complex projection coefficients $C_{1,2}$
{[}see Eq.~\eqref{eq:c12}{]} for RH (solid) and LH (dashed) circularly
polarized incident wave. (d) Analytically calculated difference in
transmission for LH/RH-polarized incident wave. Here $a_{1}=13$ mm,
$a_{2}=10$ mm, $d=10$ mm, $\phi=45^{\circ}$, and $\psi=0^{\circ}$
\label{fig:dichroism}}
\end{figure}

The dielectric permittivity tensor \eqref{eq:epsilon_eff} corresponds
to an absorbing nonmagnetic crystal, which in general has two distinct
eigenmodes with different polarizations, phase velocities and absorption
coefficients \cite{fedorov}. The medium's eigenmodes $\mathbf{h}_{1,2}$
and their associated refractive indices $n_{1,2}$ are given by eigenvectors
and eigenvalues of the refractive index tensor \cite{barkovsky} \begin{equation}
\widehat{N}_{H}=\left[\left(-\mathbf{n}^{\times}\widehat{\varepsilon}_{\text{eff}}^{-1}\mathbf{n}^{\times}\right)^{-}\right]^{-1/2}.\label{eq:refr_tensor}\end{equation}
Here $\mathbf{n}$ is an unit vector of the wave normal, the operator
$\mathbf{n}^{\times}$ is defined as $(\mathbf{n}^{\times})\mathbf{u}=\left[\mathbf{n}\times\mathbf{u}\right]$,
and $\widehat{A}^{-}$ denotes a pseudoinverse tensor to $\widehat{A}$.

In Fig.~\ref{fig:dichroism}a the real and imaginary parts of $n_{1,2}$
%
{}are shown for propagation direction orthogonal to the meta-atoms plane.
One can clearly see that absorption bands for the two eigenmodes are
different, which is characteristic for dichroic media. Fig.~\ref{fig:dichroism}b
shows the transmission and absorption spectra of the medium given
by Eq.~\eqref{eq:epsilon_eff} for an incident wave with polarization
coincident with the eigenmodes ($\mathbf{h}_{1,2}$). We calculate
the transmission spectrum from $\widehat{\varepsilon}_{\text{eff}}$
using generalized transfer matrix techniques (see, e.g., \cite{6-Borzdov97}).
Due to the small variation in the real part of the refractive indices,
reflection from such an effective medium slab is fairly small, and
the dips seen in the transmission spectra are primarily due to absorption. 

For an arbitrarily polarized incident wave, the slab acts as an absorbing
polarization filter splitting the incident field $\mathbf{H}$ into
two waves with polarizations parallel to the crystal eigenvectors
$\mathbf{h}_{1,2}$, namely \begin{equation}
\mathbf{H}=C_{1}\mathbf{h}_{1}+C_{2}\mathbf{h}_{2},\label{eq:c12}\end{equation}
where $C_{1,2}$ are complex projection coefficients. The transmission
of the effective medium is then determined by the relations (i) between
complex projection coefficients $C_{1,2}$ and (ii) between absorption
coefficients for the eigenwaves.

\begin{figure*}[t]
\begin{centering}
\includegraphics[width=1\textwidth]{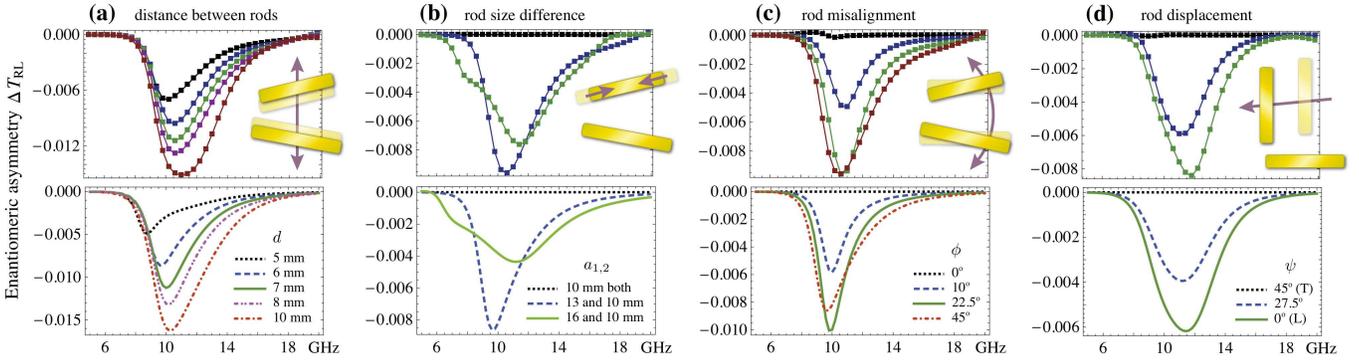}
\par\end{centering}

\caption{(Color online) Numerical (top) and analytical (bottom) dependence
of $\Delta T_{RL}$ on (a) inter-rod distance $d$, (b) difference
in rod length $a_{1}-a_{2}$, (c) rod misalignment angle $\phi$,
and (c) rod displacement angle $\psi$ for $\phi=90^{\circ}$ (as
the dimer changes between T- and L-shaped).\textbf{ }Unless specified
otherwise, $a_{1}=13$ mm, $a_{2}=10$ mm, $d=6$ mm, $\phi=45^{\circ}$,
and $\psi=0^{\circ}$. \label{fig:results}}
\end{figure*}

Fig~\ref{fig:dichroism}c shows the projection coefficients $C_{1,2}$
for circularly polarized incident waves. In this case, it can be seen
that $|C_{1}|=|C_{2}|$. However, in the spectral range of strong
absorption the coupling of RH and LH incident waves to the crystal
eigenmodes $\mathbf{h}_{1,2}$ is very different ($C_{1,2}^{R}\neq C_{1,2}^{L}$),
leading to planar chirality in the form of enantiomeric asymmetry
in the transmisison $\Delta T_{RL}=T_{R}-T_{L}$ (Fig.~\ref{fig:dichroism}d). 

Such a difference can be used to quantify the \textquotedblleft{}strength\textquotedblright{}
of planar chiral properties in a particular structure. Fig.~\ref{fig:results}
shows the spectra $\Delta T_{RL}$ for various shapes of the dimer,
analytical results from Eqs.~\eqref{eq:eq1}--\eqref{eq:eq4} compared
to the results of direct 3D frequency-domain numerical simulations
\cite{hfss}. It can be seen that the effective medium model offers
a good coincidence with numerical results.

Most rod dimers are seen to exhibit chiral properties unless the dimer
has an in-plane mirror symmetry, which makes the 2D enantiomers indistinguishable
and therefore enforces $\Delta T_{RL}=0$. For the geometrical tansformation
considered, such symmetry is achieved in the following cases: (i)
for rods of equal length (V-shaped dimer, Fig.~\ref{fig:results}b);
(ii) for parallel rods (II-shaped dimer, Fig.~\ref{fig:results}c);
(iii) for a T-shaped dimer (Fig.~\ref{fig:results}d). In all these
cases, the analytical model correctly predicts the absence of enantiomeric
asymmetry. Otherwise, chiral properties appear to be stronger when
enantiomers are most distinct, so there is an optimum rod misalignment
angle $\phi\simeq22.5^{\circ}$ in Fig.~\ref{fig:results}c. It is
also necessary that the resonances of the individual rods have some
degree of spectral overlap. Hence, $\Delta T_{RL}$ depends on the
length mismatch between the rods in a non-monotonic way: it first
increases when the dimer deviates from the achiral V shape, reaches
the maximum value, and then decreases again with a pronounced resonance
splitting as $\Delta a$ becomes greater (see Fig.~\ref{fig:results}b).
Finally, $\Delta T_{RL}$ becomes smaller as the distance between
the rods decreases (Fig.~\ref{fig:results}a).

Since the coupling between meta-atoms in the 2D lattice was neglected
both in the analytical model and in the simulations, it can be concluded
that planar chirality in the PCMs under study is intrinsic (i.e.,
attributable to the geometry of the meta-atom itself) rather than
extrinsic (attributable to the meta-atom arrangement, e.g., in tilted-cross
arrays \cite{KseniaPRA09}). This agrees with earlier time-domain
simulation results \cite{7-Kremers09}. With the proposed model extended
to include the inter-atom coupling, explicit account of intrinsic
vs.~extrinsic effects in PCMs can be given. Combining the intrinsic
and extrinsic contribution in the same PCM can be used to maximize
its chiral properties. 

In conclusion, we have shown that plasmonic rod dimers act as elementary
planar chiral meta-atoms. Using a simple coupled-dipole analytical
approach, we show the presence of elliptical dichroism and enantiomeric
asymmery. We have explored the chiral properties of rod dimers of
different geometries and demonstrated that symmetric dimers are achiral,
whereas the structures with most distinct 2D enantiomers have the
most pronounced chiral properties. Analytical results are in good
agreement with direct frequency-domain numerical simulations. 

This work was supported in part by the Deutsche Forschungsgemeinschaft
(FOR 557) and the Natural Sciences and Engineering Research Council
of Canada (NSERC).

\end{document}